# Determination of the intrinsic ferroelectric polarization in orthorhombic HoMnO$_3$

S M Feng[1,*], Y S Chai[2,*], J L Zhu[1], N Manivannan[2], Y S Oh[2], L J Wang[1], Y S Yang[2], C Q Jin[1,3], and Kee Hoon Kim[2,3]

[1]*Institute of Physics, Chinese Academy of Sciences, P.O. Box 603, Beijing 100080, People's Republic of China*

[2]*CENSCMR, Department of Physics and Astronomy, Seoul National University, Seoul 151-742, Republic of Korea*

E-mail: khkim@phya.snu.ac.kr and JIN@aphy.iphy.ac.cn

## Abstract

Whether large ferroelectric polarization $P$ exists in the orthorhombic HoMnO$_3$ with the $E$-type antiferromagnetic spin ordering or not remains as one of unresolved, challenging issues in the physics of multiferroics. The issue is closely linked to an intriguing experimental difficulty for determining $P$ of polycrystalline specimens that conventional pyroelectric current measurements performed after a poling procedure under high dc electric fields are subject to large errors due to the problems caused by leakage currents or space charges. To overcome the difficulty, we employed the PUND method, which uses successively the two positive and two negative electrical pulses, to directly measure electrical hysteresis loops in several polycrystalline HoMnO$_3$ specimens below their Néel temperatures. We found that all the HoMnO$_3$ samples had similar remnant polarization $P_r$ values at each temperature, regardless of their variations in resistivity, dielectric constant, and pyroelectric current levels. Moreover, $P_r$ of ~0.07 µC/cm$^2$ at 6 K is consistent with the $P$ value obtained from the pyroelectric current measurement performed after a short pulse poling. Our findings suggest that intrinsic $P$ of polycrystalline HoMnO$_3$ can be determined through the PUND method and $P$ at 0 K may reach ~0.24 µC/cm$^2$ in a single crystalline specimen.

---

[*] These authors equally contributed to this work.

PCAS numbers: 61.50.Ks, 75.85.+t, 75.50.Ee, 77.80.-e

# 1. Introduction

In recent years, great attention has been paid to multiferroic materials, inside which several ferroic orders such as (anti-)ferromagnetism and (anti-)ferroelectricity can coexist and couple with each other [1]. Multiferroic materials not only possess eminent application potential as next-generation devices, but also provide scientific motivations to understand the unprecedented, gigantic cross-coupling effects among multiple order parameters in solids [2]–[6]. Of particular interest in this study are improper magnetic ferroelectrics, in which ferroelectricity is induced by the primary magnetic order parameter so that it can be sensitively controlled by the application of magnetic fields [4].

Several theoretical models have been proposed to explain the inversion symmetry breaking in the magnetic ferroelectrics [4]. (1) The spiral spin ordering, often realized in frustrated spin systems, can induce ferroelectric polarization ($P$) via inverse process of Dzyaloshinskii-Moriya spin-spin interaction that is sensitive to spin-orbit coupling. Either a spin-current model [5] or free-energy consideration [6, 7] can describe the phenomena effectively and have explained ferroelectricity observed in various multiferroics, e.g. $Ni_3V_2O_8$ [8], $TbMnO_3$ [9], and $CoCr_2O_4$ [10, 11]. (2) Experimental studies on $RMn_2O_5$ ($R$ = Tb and Bi) [12]–[14], and more recently on $Ca_3CoMnO_6$ [15] have shown that exchange striction in a frustrated spin network with asymmetric charge distribution can result in a finite $P$, even in collinear spin-ordered patterns. This process involves a magneto-elastic lattice modulation and subsequently appearance of a generally small, uncompensated $P$. Most of the currently known magnetic ferroelectrics seem to be categorized into one of these two cases while both mechanisms can be valid in some materials, as recently suggested in $RMn_2O_5$ ($R$ = Y, Ho and

Tm) [16]–[18].

Although successes in understanding those mechanisms and finding new magnetic ferroelectrics have driven recent progresses, their small $P$ (~ 0.1 µC/cm$^2$ or less) constitutes a serious drawback for application. This is partly because the spin-orbit coupling is fundamentally small in solids or $P$ is almost compensated in improper ferroelectrics. On the other hand, a series of recent theoretical reports [19]–[21] have suggested a new mechanism to generate $P$ as large as 6 µC/cm$^2$ in orthorhombic perovskite HoMnO$_3$ ($o$-HoMnO$_3$), which has a collinear $E$-type antiferromagnetic (AFM) structure with ferromagnetic (FM) zigzag spin chains. In this special spin configuration, $P$ is thought to arise from a gain in the band energy owing to the hopping of $e_g$-electrons along the FM chain, which overcomes the competing elastic energy. Although, this can be classified as case (2) above in a broad sense, the unique possibility of having large $P$ in magnetic ferroelectrics is important for future application of multiferroics and thus has received particular attention.

However, the confirmation or checking of above theoretical predictions still remains as one of the long standing experimental challenges. First of all, only limited experimental results are available. An early experimental attempt only found $P$ ~ 0.008 µC/cm$^2$ for $o$-HoMnO$_3$ through measurements of pyroelectric currents [22], which is much smaller than the expected $P$ ~ 6 µC/cm$^2$ [19]–[21]. As $o$-HoMnO$_3$ should be synthesized under high pressure, it exists in a polycrystalline form so that reliable determination of $P$ becomes intriguingly difficult; the most common method of using the pyroelectric current measurements is subject to significant errors in the case of polycrystalline ferroelectrics. This is due to imperfect sample quality such as porosity, oxygen deficiency, or large leakage current, which result in incomplete electric poling and space charge effects [23]. The space charge refers to the surface charges that can be trapped in the grain boundary of polycrystals or the ferroelectric domain boundary during an electric poling procedure. Depending on whether the electric fields are applied above or below a ferroelectric ordering temperature, the space charges are called as the $P_2$ or $-P_2$ type [23], respectively, and these two types can be a source of erroneous charges that may interfere with intrinsic pyroelectric currents. An alternative method for determining $P$ is the direct measurements of the *P-E* hysteresis loop using the Sawyer-Tower circuit [24]. However, even this method is well known to be unreliable when relatively large leakage currents exist to result in highly distorted hysteresis curves without showing saturation of *P*.

In this study, we attempt to determine the intrinsic *P* of *o*-HoMnO$_3$ by employing a commonly well established technique of the *P-E* loop measurement i.e. the PUND (Positive-Up Negative-Down) technique [25] that greatly overcomes the experimental difficulties existing in those previous methods. By using high-pressure techniques, we synthesized high-quality *o*-HoMnO$_3$ polycrystalline samples under various annealing conditions. Even though there exist variations in the resistivity, dielectric loss, and *P* values from pyroelectric current measurements among prepared samples, we obtained a similarly large *P* of ~ 0.07 μC/cm$^2$ near 6 K in all the *o*-HoMnO$_3$ samples investigated by the PUND method. Our observation clearly suggests that the intrinsic polarization of a single crystalline *o*-HoMnO$_3$ could be over ~ 0.24 μC/cm$^2$ upon extrapolation to the zero temperature limit.

**2. Experimental details**

Single-phase hexagonal HoMnO$_3$ was synthesized by the standard solid-state reaction method in air. Stoichiometric amounts of high-purity oxides of $Ho_2O_3$ and $Mn_2O_3$ were mixed, ground thoroughly, and sintered sequentially at 900, 1100, and 1350 °C with intermediate grinding until a single hexagonal phase was obtained [26]. The hexagonal precursor was transformed into an orthorhombic structure with space group *Pbnm* by high-pressure sintering at 5 GPa and 900 °C for 30 min. The as-grown samples were further post-annealed at 350 °C for 24 h in $O_2$ and $N_2$ atmospheres to obtain the $O_2$-annealed and $N_2$-annealed *o*-HoMnO$_3$ samples, respectively.

X-ray diffraction (XRD) measurements were performed for all the samples at room temperature with a high-power diffractometer using Cu $K_\alpha$ radiation (M18AHF, MAC SCIENCE, Japan). No impurity phases were detected in the XRD pattern of any of the samples investigated. The structural parameters were refined by the Rietveld method using the GSAS program. As shown in figure 1(a), the refinement results for the as-grown *o*-HoMnO$_3$ are in good agreement with the orthorhombic structure (space group *Pbnm*). The obtained lattice parameters $a = 5.24857(4)$ Å, $b = 5.83992(4)$ Å, and $c = 7.35685(5)$ Å are consistent with those reported in [27] and [28]. For the $O_2$-annealed ($N_2$-annealed) sample, both *b* and *c* become larger (smaller) than those of the as-grown sample while *a* stays almost constant, as shown in figure 1(b). With the above lattice parameters, the unit cell volumes can be calculated as 225.277, 225.926 and 226.782 Å$^3$ for $N_2$-annealed, as-grown and $O_2$-annealed sample, respectively. The increase (decrease) of the unit cell volume after $O_2$ ($N_2$) annealing implies a change in the oxygen content of the sample. The electrical and magnetic properties of the samples were measured at various temperatures (*T*) and magnetic fields by using of a physical property measurement system (PPMS, Quantum Design). Magnetization data were measured with a vibrating sample magnetometer attached to the PPMS.

To reduce the effect of trapped space charges near the grain boundaries of a polycrystalline specimen, it can be quite important to measure the density of the specimen. We found that the density of the as-grown sample was 95.8 % and does not change significantly with the additional annealing procedures. For electrical measurements, the samples were polished into thin plates with a typical thickness of 0.3 mm and an area of 2 mm$^2$. The dielectric constant $\varepsilon$ and the loss tan$\delta$ were measured using an Agilent 4284A LCR meter and an Andeen Hagerling 2550A capacitance bridge at a frequency of 1 kHz. Ferroelectric polarization was determined in two different ways, i.e., pyroelectric current $J_p$ and *P-E* hysteresis loop measurements based on the PUND method [25]. A recent literature that applied the PUND method to the multiferroic materials also termed it as the double-wave-method [29]. For the $J_p$ measurements, the samples were poled at 50 K in a dc electric field $E_{poling}$, and then cooled down to 5 K in the electric field. After shorting the electrodes at 5 K for 15 min, $J_p$ was measured upon warming in zero electric field. *T*-dependent polarization $P(T)$ was calculated by integrating $J_p$. During the poling procedure, however, the required large $E_{poling}$ could not be fully applied because of the relatively small resistivity of the samples at around 50 K. Therefore, as will be shown below, the $P(T)$ value at 5 K did not reach saturation for any sample, even after applying the highest possible $E_{poling}$ before breakdown. From the dielectric constant and the loss measurements, the resistivity values at 50 K and 1 kHz of the as-grown, $O_2$-, and $N_2$-annealed samples were estimated to be $1.2 \times 10^8$, $1.5 \times 10^8$, and $3 \times 10^8$ ($\Omega$ cm), respectively. The *P-E* loop measurements based on the PUND method are well described in [25] and [29]. In a conventional *P-E* hysteresis measurement using a Sawyer-Tower circuit, the shape of the hysteresis loop is determined not only by hysteretic $P$ variations but also by resistance/capacitance changes in the ac electric field. When the specimen investigated has particularly large leakage, the shape of the $P$-$E$ loop is severely distorted, resulting in often erroneous estimation of the intrinsic $P$. On the other hand, in the PUND or double-wave-method, only the hysteretic $P$ component can be extracted by applying a series of voltage pulses to the specimen in a Sawyer–Tower circuit. Figure 2(a) illustrates the six voltage pulses—P0, N0, P1, P2, N1, and N2—used in this method. The first two pulses are used to fully align FE domains first in the positive direction and subsequently in the negative direction. During the next two positive (negative) pulses, two independent curves of effective $P$ changes are recorded, and the two positive (negative) curves are subtracted from each other to produce a half $P$-$E$ loop for $E > 0$ ($E < 0$), as shown in figure 2(b). In this procedure, the resistive and capacitive components are subtracted out so that only the hysteretic components are supposed to appear in the $P$-$E$ loop. Therefore, the PUND method has great advantages over the conventional $P$-$E$ loop scheme, particularly for polycrystalline ceramics with relatively large leakage currents. Although our setup is based on the scheme in [25] and [29], it has been significantly modified to improve its performance: (1) the triangle wave, instead of square or sine wave, was applied as the driving signal because it can provide constant $dE/dt$ [30]; (2) a short pulse duration of approximately 0.25 ms was employed so as to reach a high $E$ without breakdown. Moreover, we waited about 5 sec between each pulse. With this modified scheme, we could avoid most of the difficulties confronted in the conventional $P$-$E$ loop as well as $J_p$ measurements.

## 3. Results

*3.1. Electrical/magnetic properties at zero magnetic fields*

Figure 3 shows the $T$-dependent dielectric constant $\varepsilon(T)$, tan$\delta$, and $P(T)$ obtained from the $J_p$ measurements for as-grown, $O_2$-, and $N_2$-annealed samples. Solid (dashed) lines represent the results in the cooling (heating) run. In our $\varepsilon(T)$ data of the three samples, several distinct anomalies, which can be associated with the known magnetic transitions of $o$-HoMnO$_3$ [27, 31] are observed. Since some of these features are not observed in the $o$-HoMnO$_3$ sample in [22], we will explain these anomalies observed in our as-grown sample as a representative case. Firstly, upon cooling, $\varepsilon(T)$ shows an enhancement at Néel temperature $T_N \approx 42$ K, at which Mn$^{3+}$ spins is known to undergo a phase transition from a paramagnetic (PM) to an incommensurate (IC) AFM structure with a sinusoidal amplitude modulation characterized by a propagation wave vector (0, $k_{Mn} \sim 0.4$, 0) [27]. Secondly, $\varepsilon(T)$ develops a peak near the the lock-in transition temperature $T_L$, at which the IC AFM phase is known to change into a commensurate $E$-type AFM structure ($k_{Mn} \sim 1/2$) [27]. The peak positions for the cooling and heating runs show clear hysteresis, due to the first-order nature of the transition. We find that the peak position for each sample in the heating run is quite close to the temperature where $P(T)$ starts to develop in figure 3(c). This is consistent with the expectation that the $E$-type AFM structure would develop a finite $P$ below $T_L$. Therefore, in this study, we attribute $T_L$ to the temperature where $\varepsilon(T)$ shows a peak. Thirdly, there exists a small kink in $\varepsilon(T)$ at around $T_3 = 8$ K, which coincides with the magnetic susceptibility peak under the magnetic field $\mu_0 H = 0.1$ tesla (figure 5(a)). It is known that, below $T_3$, the Ho$^{3+}$ spins have a canted ground state, in which the angle between the Ho$^{3+}$ magnetic moment and the $a$-axis becomes approximately

60 degree [27]. Interestingly, tanδ in figure 3(b) shows a prominent sharp peak at $T_3$ so that the $T_3$ position can be more easily identified from this peak. Lastly, we note that additional broad peaks exist just below $T_L$ in the $\varepsilon(T)$ heating curve, particularly at around 16 K for the as-grown sample and at around 20 K for the $N_2$-annealed sample. With the current data set alone, it is not easy to conclude whether this feature can be associated with intrinsic magnetic transitions such as a collinear to non-collinear transition of the $Ho^{3+}$ spins or extrinsic effects possibly existing in the polycrystalline sample, such as trapped charges. However, the overlap of the broad feature in $\varepsilon(T)$ with the lock-in transition in the cooling curve is mostly likely to result in the anomalously broader $\varepsilon(T)$ peak than that in the heating curve.

From all the features seen in $\varepsilon(T)$ at zero-tesla, the expected magnetic transitions for each sample are summarized in the inset of figure 3(a). The location of $T_L$ is noticeably varied by the annealing treatments; $T_L$ decreases with enhanced hysteresis in the order of $N_2$-annealed, as-grown, and $O_2$-annealed samples. Moreover, $T_3$ slightly increases in this order while $T_N$ is almost constant. It is well known that ceramic oxide samples prepared under high-pressure and high-$T$ can contain oxygen vacancies and internal stress. Post annealing in an $O_2$ ($N_2$) atmosphere at around 350 °C is expected to release the internal stress and increase (decrease) the oxygen content. Since the $E$-type AFM spin ordering at $T_L$ is accompanied by the first-order structural transition involving changes in the Mn-O-Mn bond angle/Mn-O bond length, its location as well as its hysteresis behavior seem to be more sensitive to internal stress. Moreover, since samples with an increased $T_L$ tend to have a higher oxygen vacancies, the structural change resulting from the presence of oxygen vacancies might be similar to the case of having smaller rare-earth ions [32], in which an increase in $T_L$ and a simultaneous decrease in the unit cell volume have been observed. Indeed, we also confirmed such correlation through the structural refinement data in figure 1(b) that the $N_2$-annealed ($O_2$-annealed) sample has smaller (larger) unit cell volume, 225.277 Å$^3$ (226.782 Å$^3$) than that of the as-grown sample, 225.926 Å$^3$ by approximately 0.3% (0.3 %). This indicates that oxygen vacancy plays a key role in changing $T_L$ via structural modification.

*3.2. Determination of intrinsic electric polarization via the PUND method*

To investigate the ferroelectric properties of all the samples, we first performed the $J_p$ measurements to obtain the $T$-dependent polarization $P(T)$, as shown in figure 3(c). To our surprise, while all the three samples show more or less similar magnetic transitions, the $P(T)$ values vary significantly over the samples; for example, $P$(5K) obtained with the maximum $E_{poling}$ are 0.0142 µC/cm$^2$, 0.0600 µC/cm$^2$, and 0.0639 µC/cm$^2$ for the as-grown, $N_2$-, and $O_2$-annealed samples, respectively. These $P$ values are much larger than the reported one (~ 0.008 µC/cm$^2$) for an $o$-HoMnO$_3$ polycrystalline sample [22]. However, large variation of $P$(5K) over the samples still casts doubt on its validity as the intrinsic $P$. The continuous increase in $P$(5K) with $E_{poling}$, as shown in the inset of figure 3(c), further supports that the maximum $E_{poling}$ before the sample breakdown was not enough to fully align the ferroelectric domains. Moreover, the maximum $E_{poling}$ applied for different samples differed significantly; 0.51 MV/m for the as-grown and 1.78 MV/m for the $N_2$-annealed sample.

To estimate the intrinsic $P$ values of these samples more reliably, we measured directly the $P$-$E$ hysteresis loop by the PUND method. Figure 4(a) summarizes thus obtained hysteresis curves for the $N_2$-annealed sample at various $T$. Indeed, the other samples showed quite similar hysteresis curves. As seen in figure 4(a), $P$-$E$ hysteresis loop shows good saturation approximately above 10 K, as the coercive electric fields are clearly smaller than 3 MV/m. As we approach the lower $T$ below 10 K, the coercive electric fields seem to increase beyond 3 MV/m, but are still much smaller than the maximum applied $E$ of 6.7 MV/m. Therefore, it is likely that remnant polarization $P_r$ at 6 K is close to the intrinsic value expected for this sample.

For all the samples, we have summarized the $P_r$ vs. $T$ in figure 4(b). Contrary to the $P(T)$ curves in figure 3(c), the $P_r$ values of all the three samples are similar at temperatures below $T_L$. At the lowest measured $T$, i.e., 6–7 K, all the three samples showed nearly saturated loops, as seen in the $N_2$-annealed sample, and the $P_r$ values are commonly close to 0.07 $\mu C/cm^2$. This observation clearly supports that all the samples have very similar intrinsic $P_r$ within the experimental error of ~ 5%, regardless of the annealing treatment.

In order to verify the large $P_r$ value obtained by the PUND method, after completing one $P$-$E$ loop at 6 K for the $N_2$-annealed sample, we increased the temperature to above $T_L$ to measure and integrate $J_p$ to obtain $T$-dependent $P$, termed as the pulse-poled polarization $P_{pls}(T)$. It is plotted in figure 4(b) for comparison. This $P_{pls}(T)$ curve closely fits to the $T$-dependent behavior of $P_r$, again supporting that the obtained $P_r$ is close to the intrinsic value of this compound. We note that the shape of the $P_r(T)$ and $P_{pls}(T)$ curves in figure 4(b) are similar to that of $o$-YMnO$_3$, but not to that of $o$-HoMnO$_3$ sample in [22], in the sense that they all have a concave (positive) curvature. Moreover, the convex (negative) curvature of $P(T)$ seen at around 15 K does not exist in the $P_r(T)$ or $P_{pls}(T)$ curves, indicating that it might not be an intrinsic property related to the spin ordering of rare-earth ions but some other extrinsic effects such as that caused by space charges, which might be trapped in the grain boundary after dc electric poling.

*3.3. Determination of magnetic/electric phase diagram*

To determine the phase diagram of *o*-HoMnO$_3$ under magnetic fields (*H*), we have also measured magnetization/dielectric constants up to 9 tesla. Figure 5 summarizes the *T*- and *H*-dependent $\varepsilon$, tan$\delta$, and *M* results, respectively, for the as-grown sample. The *M*/*H* curves in figure 5(a) show a clear reduction at $T_3 \sim$ 8 K owing to the canting of the Ho$^{3+}$ spins at zero-field. With the further increase in $\mu_0 H$ below 1 tesla, *M*/*H* at $T < T_3$ increases, and $T_3$ shifts to lower temperatures. At $\mu_0 H$ = 1.5 tesla, the sharp reduction in the *M*/*H* curve disappears, but a kink remains at around $T_3$ = 6 K, indicating that the Ho$^{3+}$ moments are aligned along the *H* direction, but only partially. At this $T_3$ position, both $\varepsilon$ and tan$\delta$ show anomalous kink and peak, respectively, as explained above in the case of $\mu_0 H$ at zero-tesla (figure 3(b)). We could confirm that the temperatures at which these electrical signatures appear decrease under finite *H* (figures 5(b) and (c)), consistent with the change in $T_3$ indicated in the *M*/*H* curves. Contrary to $T_3$, $T_L$ slightly increased under *H*, as confirmed by the peak shift in figure 5(b).

On the other hand, the sudden jump-like features seen in the *H*-dependent *M* and $\varepsilon$ curves, and the hump in the tan$\delta$ curve in figures 5(d)-(f) represent a metamagnetic transition at $\mu_0 H_m \approx$ 1 tesla [27, 33]. This metamagnetic transition is observed below $T_3$. Therefore, it is attributed to the *H*-induced rearrangement of the canted AFM state of Ho$^{3+}$ ($4f^{10}$, $^5 I_8$) spins to a partially aligned spin state because the saturated moment of Ho$^{3+}$ is expected to be 10 $\mu_B$. At $T \geq$ 10 K, the *M*(*H*) curves show a rather monotonous increase with a small positive curvature at least up to 30 K, resembling the shape of the Brillouin function. At 50 K above $T_N$, the $M(H)$ curve is almost linear. Consistent with this, $\varepsilon(H)$ also shows a monotonous decrease with a negative curvature below 20 K, in which the $E$-type collinear spin state is stabilized. At 30 K, where the IC-AFM state of $Mn^{3+}$ spins is stabilized, $\varepsilon(H)$ slightly increases up to 9 tesla, similar to the case of the IC-AFM phase in $TbMnO_3$ [9]. At $T = 50$ K, at which the PM state is stabilized, $\varepsilon(H)$ is almost independent of $H$. All these observations show that the magnetodielectric effects of $o$-$HoMnO_3$ are closely coupled to the spin states of $Mn^{3+}$ and $Ho^{3+}$ ions.

The magnetic/electric phase diagram is constructed in figure 6, based on the anomalies seen in the data in figure 5. We note that the three trajectories of $T_N$, $T_L$, and $T_3$ under $H$ seen in our phase diagram are quite consistent with those seen in a magnetic phase diagram constructed from the neutron diffraction study in [27]. Although a trace of the metamagnetic transition field $H_m$ has not been drawn in [27], the reported $H_m$ value is also consistent with our phase diagram. Finally, we have studied similar dielectric/magnetic properties in $\mu_0 H$ up to 9 tesla for the $O_2$- and $N_2$-annealed samples, confirming that their phase diagrams are qualitatively similar to that of the as-grown sample in figure 6.

## 4. Discussion

It is noteworthy that the overall shape of the three $\varepsilon(T)$ curves in figure 3(a) is rather similar to that of the reported $o$-$YMnO_3$, but quite different from that of $o$-$HoMnO_3$ in [33]. Even the lattice parameters of our as-grown $o$-$HoMnO_3$ are indeed somewhat close to those published for $o$-$YMnO_3$ ($a = 5.25975(2)$ Å, $b = 5.83535(2)$ Å, and $c = 7.35568(3)$ Å) [34], which is reasonable because the ionic radius of $Y^{3+}$ (= 1.019 Å) is close to that of $Ho^{3+}$ (= 1.015 Å)

[35]. Therefore, the qualitative similarity between the $\varepsilon(T)$ curve of our as-grown $o$-HoMnO$_3$ and that of $o$-YMnO$_3$ might stem from the similarity in the microscopic structural parameters. Besides the similarity, we also note that the $\varepsilon(T)$ curve of our as-grown $o$-HoMnO$_3$ sample clearly shows additional dielectric anomalies that are closely correlated with the magnetic transitions related to Ho$^{3+}$ ions. This observation supports that our as-grown sample is indeed a high quality one although its resistivity was a little lower than that of the annealed samples. Moreover, the existence of similar magnetic/electric phase boundaries in O$_2$- and N$_2$-annealed samples implies that all the three samples are close to the optimal sample, except for the variations in their dielectric losses.

We note in figure 3(b) and the inset of figure 3 (c) that the dielectric loss, i.e. tan$\delta$ at 50 K is inversely proportional to the maximum value of $E_{\text{poling}}$, $E_{\text{poling}}^{\text{max}}$ that can be applied without breakdown. As the dc electric field for the $J_\text{p}$ measurements is applied at 50 K, the conductivity of each specimen, proportional to the dielectric loss at 50 K, will be a decisive factor for determining $E_{\text{poling}}^{\text{max}}$. In other words, a high conductivity at 50 K is likely to reduce $E_{\text{poling}}^{\text{max}}$. This also suggests that the small $P$(T) obtained from the as-grown sample in figure 3(a) might be the simple consequence of the imperfect poling limited by the conductivity at 50 K. In contrast, the smaller loss at 50 K in the N$_2$- and O$_2$-annealed samples allowed the larger $E_{\text{poling}}^{\text{max}}$ and consequently larger $P$(T) values. Thus, annealing process appears to be essential in lowering the dielectric loss of the present high-pressure-synthesized materials. Although this finding indicates that the $P$(5K) ~ 0.06 μC/cm$^2$ seen in the N$_2$- and O$_2$-annealed samples might be close to the intrinsic $P$ of $o$-HoMnO$_3$, the lack of saturation in the $P$(5K) even at the $E_{\text{poling}}^{\text{max}}$ for both N$_2$- and O$_2$-annealed samples (inset of figure 3(c)) cannot support this conjecture. Therefore, our results imply that the $P(T)$ values estimated from the $J_p$ measurements in figure 3(c) as well as the one published in [22], should not be attributed to the intrinsic $P$ of $o$-HoMnO$_3$.

On the other hand, we observe very similar $P_r$ values for all the three samples from the PUND method, indicating that the $P_r$ value is close to the intrinsic one of $o$-HoMnO$_3$. Although there exist some changes in the oxygen content and/or internal stress over the three samples investigated here, it is likely that the PUND method can extract the intrinsic $P_r$ of $o$-HoMnO$_3$ regardless of those differences. This can be summarized as two important conclusions: (1) our as-grown sample is already close to the high quality sample and (2) variations in the oxygen vacancy and/or stress under the specific heat treatment conditions we chose enhance the resistivity and slightly change the lattice parameters but do not alter the intrinsic $P_r$ appreciably.

The $P_r$(6K) value of ~ 0.07 µC/cm$^2$ in $o$-HoMnO$_3$ and its temperature dependence suggests that $P_r$ can reach ~ 0.08 µC/cm$^2$ upon extrapolation to the zero-$T$ limit. This value is much larger than that reported previously (~ 0.008 µC/cm$^2$) by almost one order of magnitude, but is still much smaller than the theoretical prediction of 6 µC/cm$^2$ [19]–[21]. In the $E$-type AFM spin ordering, $P$ is predicted to be uniaxial along the $a$-axis. $P_r$(0K) ~ 0.08 µC/cm$^2$ in a polycrystalline specimen with a randomly oriented $a$-axis implies that a single crystal would have at least three times larger $P$ ~ 0.24 µC/cm$^2$. Although this expected $P_r$(0K) ~ 0.24 µC/cm$^2$ for a single-crystal specimen is still smaller than the theory prediction, we should stress that the value is one of the biggest among the magnetism-driven ferroelectrics. Therefore, our results call for further efforts to explain this discrepancy between the experimental and theoretical studies as well as to find a way of increasing $P$ further in the multiferroics with the $E$-type spin structure. For the experimental part, to confirm the predicted $P \sim 0.24$ μC/cm$^2$ in a single-crystal specimen would be worthwhile, while for the theoretical part, elaboration or reexamination of the existing theories might be necessary.

## 5. Conclusions

By employing direct measurements of the polarization-electric field hysteresis loop based on the PUND method, we determined the intrinsic electric polarization for several high-quality orthorhombic HoMnO$_3$ specimens synthesized under high pressure. Although the ferroelectric polarization estimated from the conventional pyroelectric current measurements were subject to large variation over the samples, all the samples showed clear electrical and magnetic transitions related to Ho$^{3+}$ and Mn$^{3+}$ spins, and commonly exhibited large polarization ~ 0.07 μC/cm$^2$ near 6 K by the PUND method. Our results show that the intrinsic ferroelectric polarization of orthorhombic HoMnO$_3$ is higher than the known in the previous experiments, but still smaller than the theoretical predictions to date.

**Acknowledgments**

This work is supported by Korean Government through National Creative Initiative Center, NRL program (M10600000238), NRF with grant numbers (2009-0083512), and by MOKE through the Fundamental R&D Program for Core Technology of Materials. Y. S. Oh was supported by Seoul R&BD (10543). This work is also supported by NSF & MOST of China through the research projects.

# Reference

[1] Spaldin N A and Fiebig M 2005 *Science* **309** 391
[2] Fiebig M 2005 *J. Phys. D-Appl. Phys.* **38** R123
[3] Ramesh R and Spaldin N A 2007 *Nat. Mater.* **6** 21
[4] Cheong S W and Mostovoy M 2007 *Nat. Mater.* **6** 13
[5] Katsura H, Nagaosa N and Balatsky A V 2005 *Phys. Rev. Lett.* **95** 057205
[6] Mostovoy M 2006 *Phys. Rev. Lett.* **96** 067601
[7] Kenzelmann M, Harris A B, Jonas S, Broholm C, Schefer J, Kim S B, Zhang C L, Cheong S W, Vajk O P and Lynn J W 2005 *Phys. Rev. Lett.* **95** 087206
[8] Lawes G, Harris A B, Kimura T, Rogado N, Cava R J, Aharony A, Entin-Wohlman O, Yildirim T, Kenzelmann M, Broholm C and Ramirez A P 2005 *Phys. Rev. Lett.* **95** 087205
[9] Kimura T, Goto T, Shintani H, Ishizaka K, Arima T and Tokura Y 2003 *Nature* **426** 55
[10] Yamasaki Y, Miyasaka S, Kaneko Y, He J -P, Arima T and Tokura Y 2006 *Phys. Rev. Lett.* **96** 207204
[11] Kim Ingyu, Oh Y S, Liu Y, Chun S H, Lee J S, Ko K T, Park J H, Chung J H and Kim K H 2009 *Appl. Phys. Lett.* **94** 042505
[12] Hur N, Park S, Sharma P A, Ahn J S, Guha S and Cheong S W 2004 *Nature* **429** 392
[13] Chapon L C, Blake G R, Gutmann M J, Park S, Hur N, Radaelli P G and Cheong S W 2004 *Phys. Rev. Lett.* **93** 177402
[14] Kim J W, Haam S Y, Oh Y S, Park S, Cheong S W, Sharma P A, Jaime M, Harrison N, Han J H, Jeon G S, Coleman P and Kim K H 2009 *Proc. Natl. Acad. Sci.* **106** 15573
[15] Choi Y J, Yi H T, Lee S, Huang Q, Kiryukhin V and Cheong S W 2008 *Phys. Rev. Lett.* **100** 047601
[16] Kim J H, Lee S H, Park S I, Kenzelmann M, Harris A B, Schefer J, Chung J H, Majkrzak C F, Takeda M, Wakimoto S, Park S Y, Cheong S W, Matsuda M, Kimura H, Noda Y and Kakurai K 2008 *Phys. Rev. B* **78** 245115
[17] Kimura H, Kamada Y, Noda Y, Kaneko K, Metoki N, Kohn K 2006 *J. Phys. Soc. Jpn.* **75** 113701
[18] Fukunaga M, Sakamoto Y, Kimura H, Noda Y, Abe N, Taniguchi K, Arima T, Wakimoto S, Takeda M, Kakurai K and Kohn K 2009 *Phys. Rev. Lett.* **103** 077204
[19] Picozzi S, Yamauchi K, Sanyal B, Sergienko I A and Dagotto E 2007 *Phys. Rev. Lett.* **99** 227201
[20] Yamauchi K, Freimuth F, Blugel S and Picozzi S 2008 *Phys. Rev. B* **78** 014403
[21] Sergienko I A, Sen C and Dagotto E 2006 *Phys. Rev. Lett.* **97** 227204
[22] Lorenz B, Wang Y Q and Chu C W 2007 *Phys. Rev. B* **76** 104405
[23] Okazaki K and Maiwa H 1992 *Jpn. J. Appl. Phys.* **31** 3113
[24] Sawyer C B and Tower C H 1930 *Phys. Rev.* **35** 269
[25] Scott J F, Kammerdiner L, Parris M, Traynor S, Ottenbacher V, Shawabkeh A and Oliver W F 1988 *J. Appl. Phys.* **64** 787
[26] Feng S M, Wang L J, Zhu J L, Li F Y, Yu R C, Jin C Q, Wang X H and Li L T 2008 *J. Appl. Phys.* **103** 026102
[27] Munoz A, Casais M T, Alonso J A, Martinez-Lope M J, Martinez J L and Fernandez-Diaz M T 2001 *Inorg. Chem.* **40** 1020

[28] Alonso J A, Martinez-Lope M J, Casais M T and Fernandez-Diaz M T 2000 *Inorg. Chem.* **39** 917

[29] Fukunaga M and Noda Y 2008 *J. Phys. Soc. Jpn.* **77** 064706

[30] Glazer A M, Groves P and Smith D T 1984 *J. Phys. E: Sci. Instrum.* **17** 95

[31] Munoz A, Alonso J A, Casais M T, Martinez-Lope M J, Martinez J L and Fernandez M T 2001 *J. Alloys Comp.* **323-324** 486

[32] Tachibana M, Shimoyama T, Kawaji H, Atake T and Takayama-Muromachi E 2007 *Phys. Rev. B* **75** 144425

[33] Lorenz B, Wang Y Q, Sun Y Y and Chu C W 2004 *Phys. Rev. B* **70** 212412

[34] Zhou J S, Goodenough J B, Gallardo-Amores J M, Moran E, Alario-Franco M A and Caudillo R 2006 *Phys. Rev. B* **74** 014422

[35] Shannon R D 1976 *Acta Crystallogr. Sect. A* **32** 751

**Figure captions**

**Figure 1.** (a) Rietveld refinement results for the XRD data of the as-grown $o$-HoMnO$_3$ sample with *Pbnm* space group. The refinement resulted in the weighted $R$-factor $R_{wp}$ = 4.47%, the profile residual factor $R_P$ = 3.24%, and $\chi^2$ = 1.952. (b) Lattice parameters for the O$_2$-annealed, as-grown, and N$_2$-annealed samples obtained from the refinement.

**Figure 2.** (a) A typical pattern of six voltage pulses used in the *P-E* hysteresis measurements by the PUND method. (b) The two *P-E* waveforms obtained from P1 and P2 (or N1 and N2) pulses are subtracted from each other to result in the half *P-E* loop.

**Figure 3.** Temperature dependence of (a) dielectric constant, (b) tanδ, and (c) *P*(*T*) from the pyroelectric current measurements for O$_2$-annealed, as-grown, and N$_2$-annealed $o$-HoMnO$_3$ samples. Inset in the panel (a) summarizes magnetic transition temperatures $T_N$, $T_L$, and $T_3$ for each sample. Inset in the panel (c) shows *P*(5K) for each sample determined from the pyroelectric current measurements after poling with different dc electric fields, $E_{poling}$, which was applied during cooling from 50 to 5 K.

**Figure 4.** (a) The *P-E* hysteresis loops determined by the PUND method for the N$_2$-annealed sample at various temperatures. (b) Temperature dependence of the remnant polarization $P_r$ for each sample and the pulse-poled polarization $P_{pls}(T)$ for the N$_2$-annealed sample.

**Figure 5.** Temperature dependent curves of (a) *M*/*H*, (b) dielectric constant, and (c) tanδ for the as-grown $o$-HoMnO$_3$ under different magnetic fields. Also shown are $H$-dependence of (d) $M$, (e) dielectric constant, and (f) tan$\delta$ for the same sample.

**Figure 6.** $T$ vs. $H$ phase diagram of the as-grown $o$-HoMnO$_3$ sample. The lines are guides for the eye.

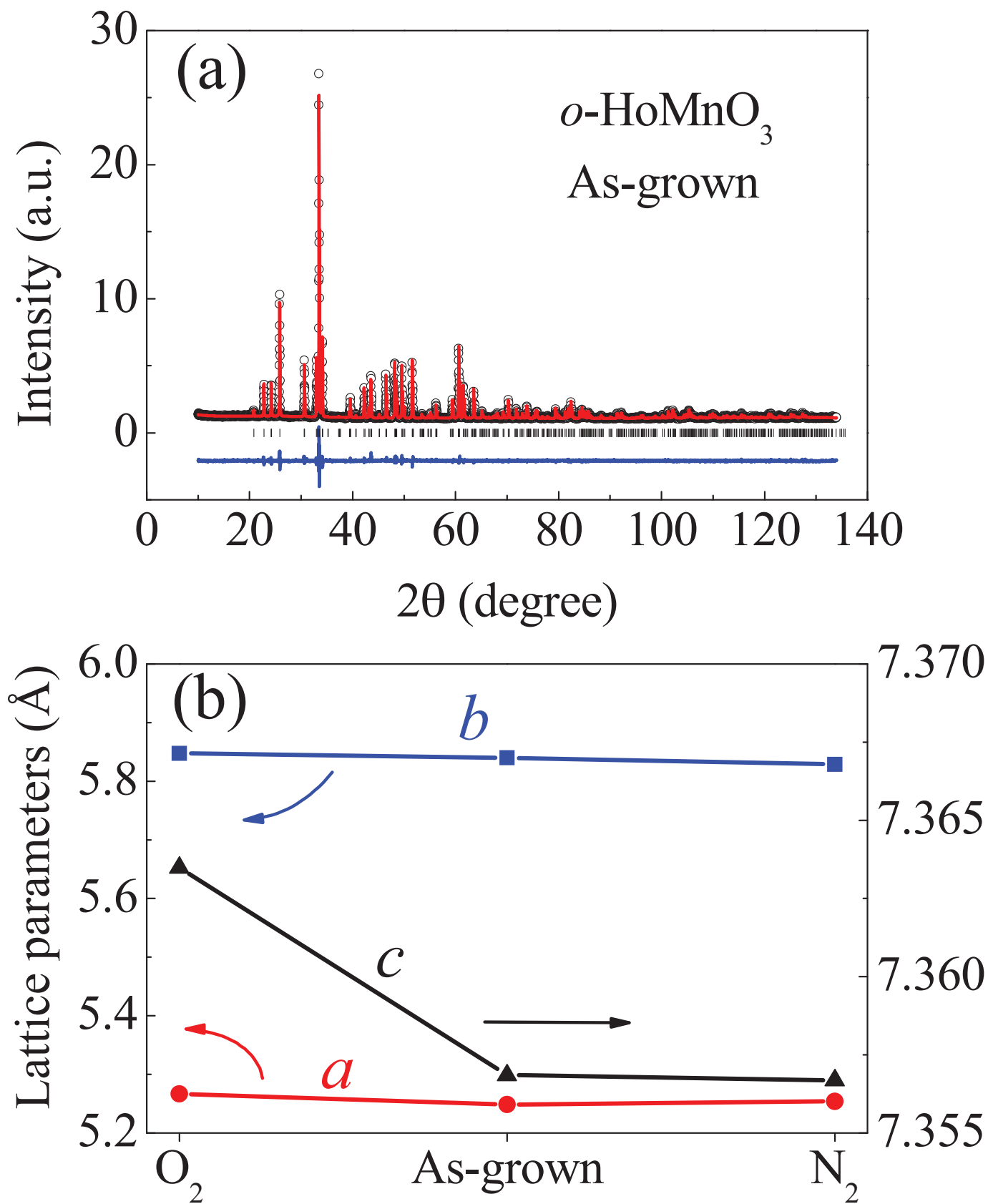

Fig. 1. S. M. Feng *et al.*

**Figure 1 (figures1.eps)**

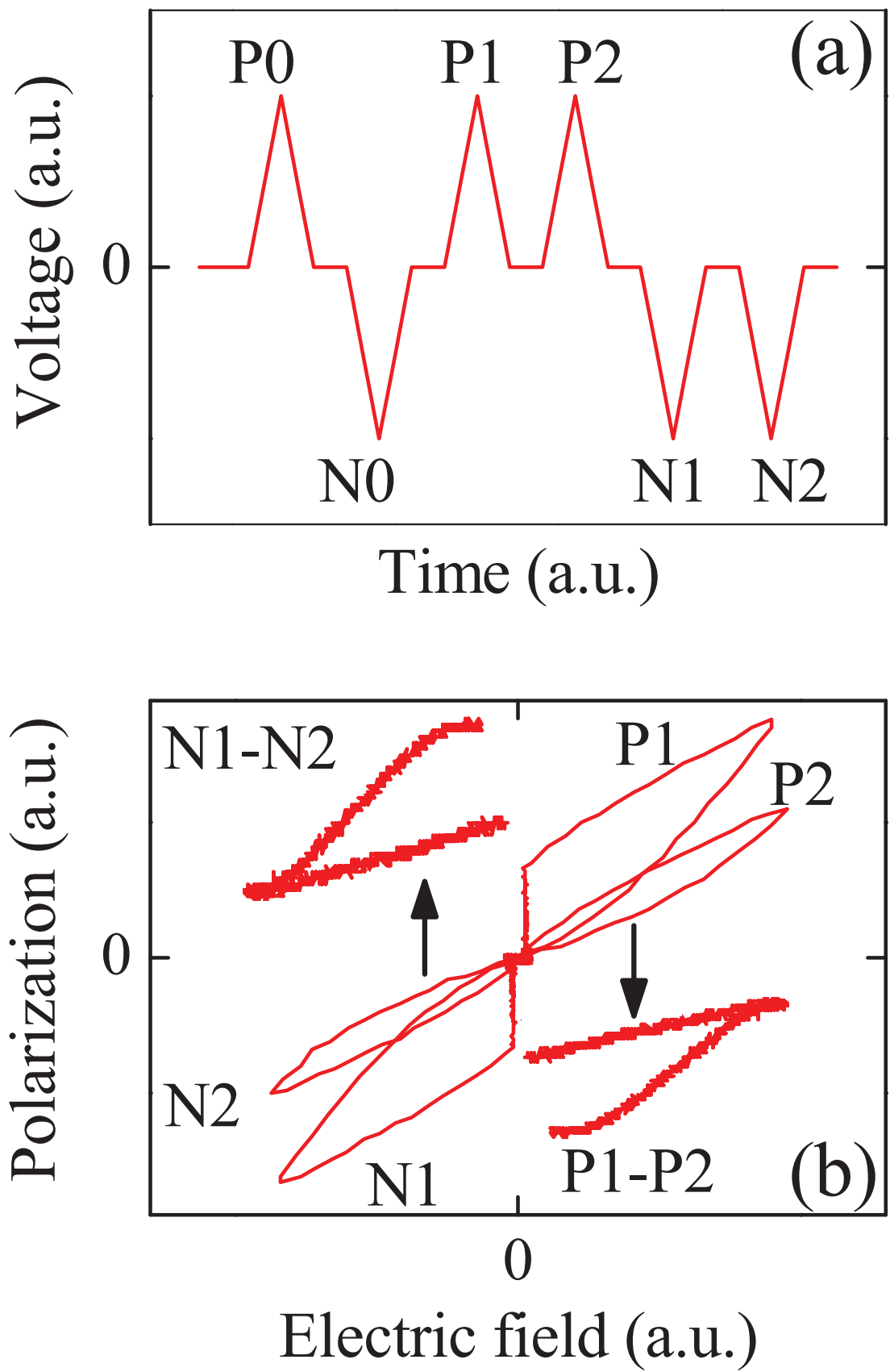

Fig. 2. S. M. Feng *et al*.

**Figure 2 (figures2.eps)**

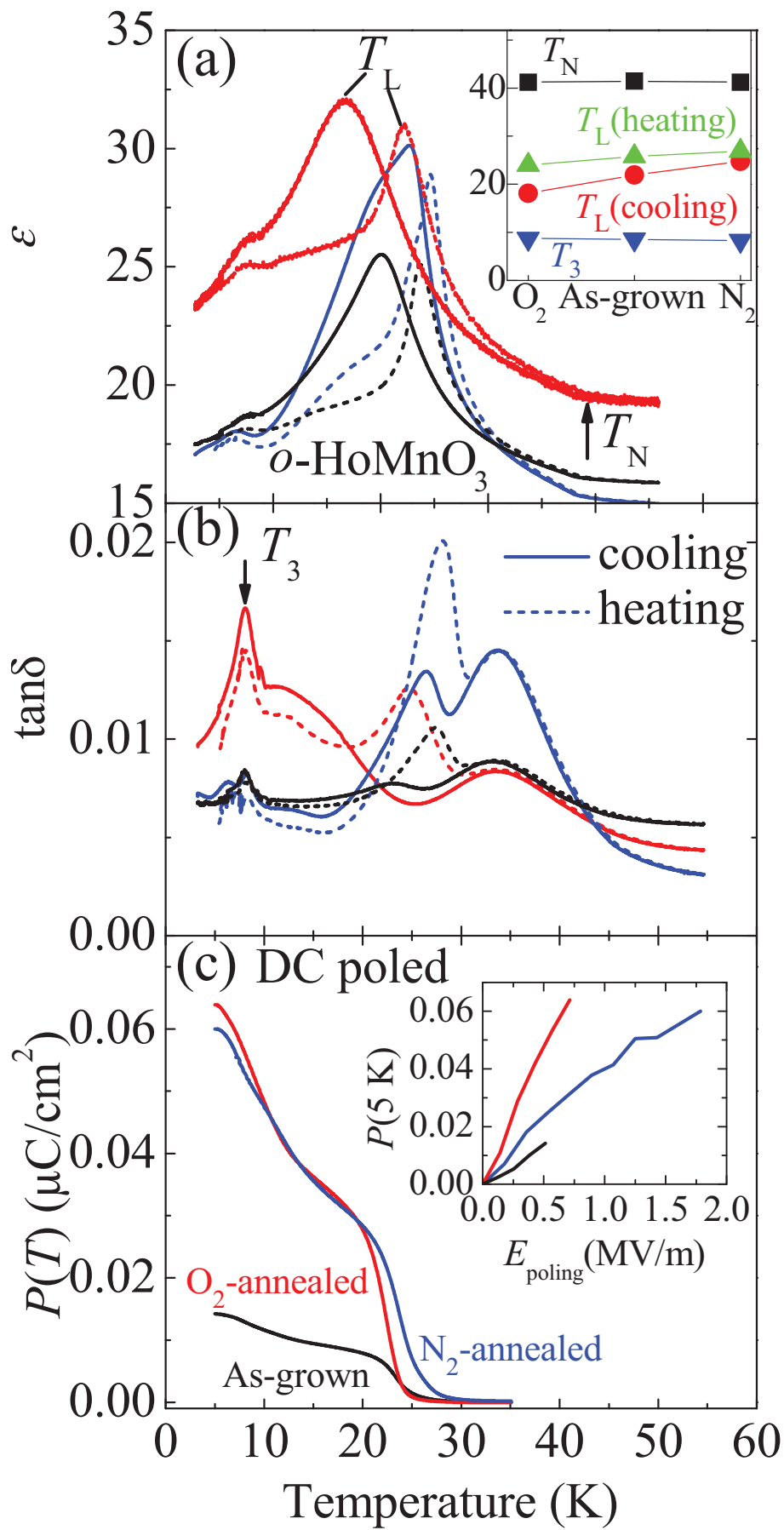

Fig. 3. S. M. Feng et al.

**Figure 3 (figures3.eps)**

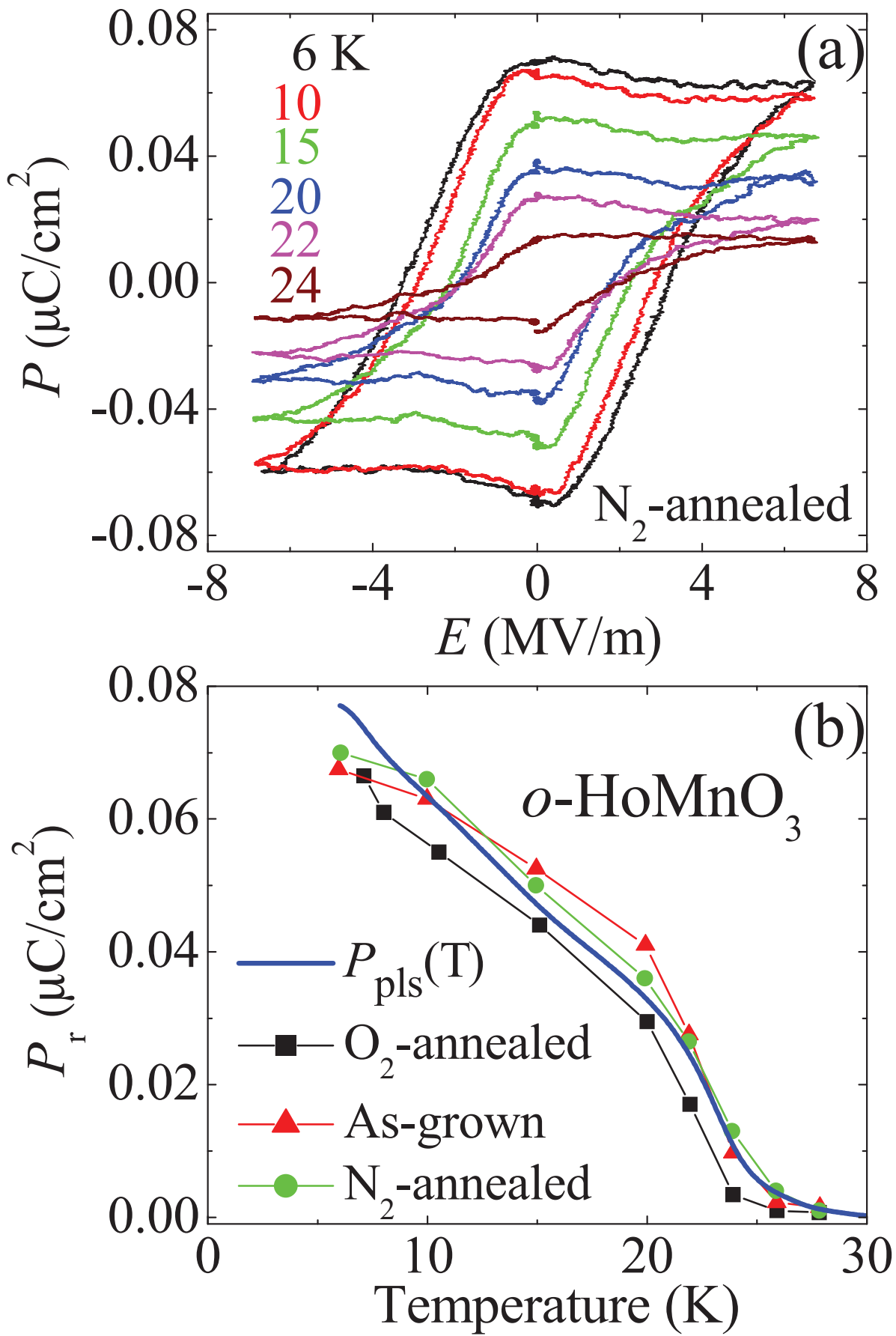

Fig. 4. S. M. Feng *et al*.

**Figure 4 (figures4.eps)**

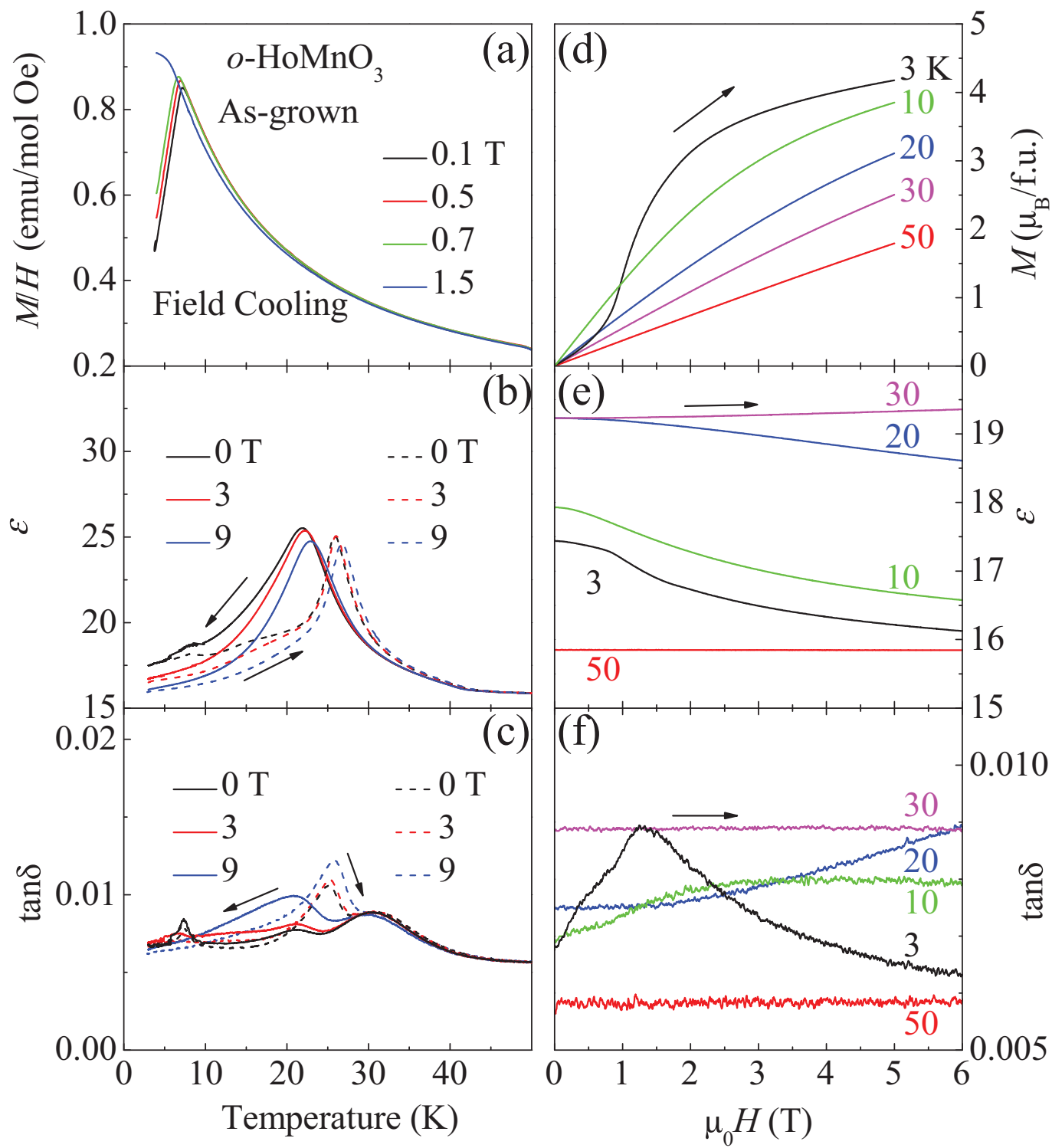

Fig.5 S.M. Feng *et al.*

**Figure 5 (figures5.eps)**

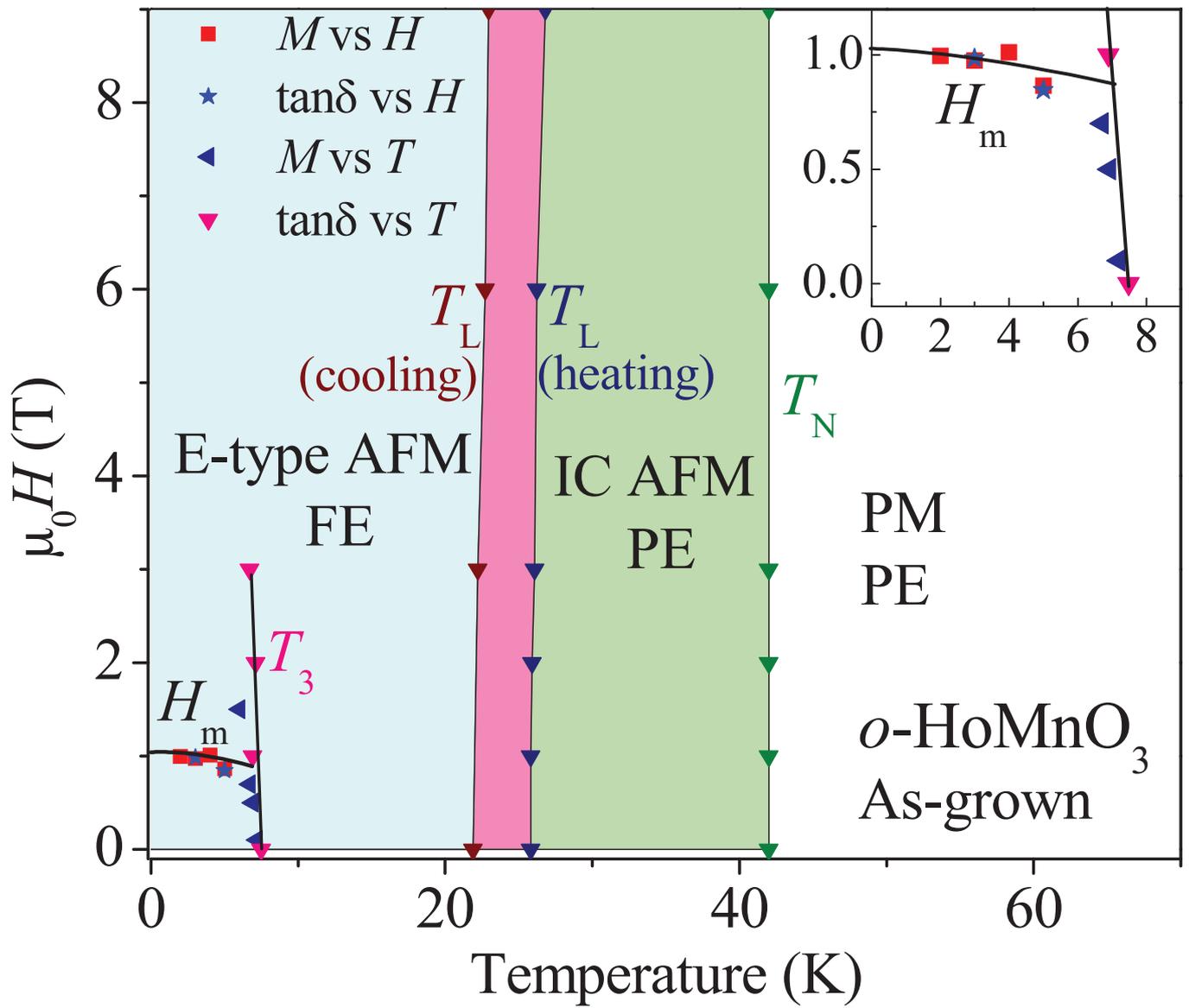

Fig. 6. S. M. Feng *et al.*

**Figure 6 (figures6.eps)**